# Aligning Enterprise Systems with the Organisation – A Sensemaking Perspective

**Petri Hallikainen**
Discipline of Business Information Systems
The University of Sydney Business School
Sydney, Australia
Email: petri.hallikainen@sydney.edu.au

**Ravi Seethamraju**
Discipline of Accounting
The University of Sydney Business School
Sydney, Australia
Email: ravi.seethamraju@sydney.edu.au

## Abstract

This paper explores the alignment of enterprise systems and organisations in the post-adoptive phase. A research model is presented focusing on the alignment decisions in the managerial search process for solutions to business problems related to enterprise systems. Based on the research model we propose research questions for future research, particularly to analyse how managerial identity would affect the alignment decisions.

## 1　Introduction

Enterprise systems are large software packages such as SAP ERP, Oracle PeopleSoft or Microsoft Dynamics NAV, which support the main business functions in organisations including human resource management, logistics, accounting, supply chain management and customer relationship management. They are expected to improve internal processes, or inter-organisational processes along the supply chain creating closer business partnerships and improve organizational performance (Davenport, 1998; Ross and Vitale, 2000; Sumner, 2000). These enterprise systems have already been in use for more than 15 years in most of the large organizations and many medium-sized organizations. These software packages come with various modules – accounting, inventory, materials management, sales, production, maintenance and others and organizations typically implement the modules depending upon their business needs, costs and capabilities. Even though huge investments are made in these systems, the features and functionality of various ES components or modules implemented in business organizations are not always fully utilised and some of them are never implemented (Botta-Genoulaz and Millet 2005).

With software vendors regularly introducing new versions with new technologies, improved features and expanded functionality, these enterprise systems require continuous upgrading to improve productivity and sustain organizational competitiveness (Hawking and Selitto 2015). Such upgrades are considered costly in terms of licence fees, downtime, implementation and learning (Demsey et al 2013). In addition, vendor charges for software upgrades and maintenance are also high and their software upgrade release cycle is considered excessive (Irani et al 2006). Driven by the software vendor's product life cycle and profit maximization goals and the risk of losing support for the existing version of the ES software, business organizations are under pressure to upgrade to the next version before they have achieved the returns on their last upgrade/implementation (Luftman et al 2013).

Discovering, learning and using the available functionality and features of ES and thereby improving efficiencies, performance and competitiveness is therefore essential for business organizations. This is not a one-off event at the time of implementation, but a continuous process. Importantly, it requires continuous alignment of organizational capabilities and priorities with the functionality and features of the enterprise systems as well as the outcomes delivered by these systems. As characterised by Orlikowski, such alignment involving technology change and adaptation in organizations is by improvisation rather than a process of managerial planning, design and intervention (Orlikowski 1996). This phenomenon is "emergent, continuous and more difficult to control and more affected by what people pay attention to than" by the designed features and functionality of the software IS/IT artefact (Weick, 1993: 348). Traditional literature, however, treats both IS and organizational components as static entities that can be controlled (Benbya and McKelvey 2006). This alignment of the IS/IT artefact and organizational components at individual, operational and strategic levels are in a continuous and dynamic interaction with one another and is not well understood in the literature. An understanding how this alignment takes place and how it facilitates the deprival of full benefits of enterprise systems in use is needed. Even though enterprise systems are in use for a number of years, research on the utilization of the potential and effects of enterprise systems in the post-adoptive stage are surprisingly few (Peng and Nunes 2009).

This research study will analyse the alignment between enterprise systems and the decisions made on the post-adoptive phase. Drawing from the sense making theory (Weick 1995) and from the theoretical model of creative search and strategic sense making by Pandza and Thorpe (2009), this study will develop a theoretical research model for the process of identifying solutions for enterprise system related business problems. The model is constructed particularly from the alignment perspective. Researching the cognitive processes and managerial actions employed for fully utilizing the potential of enterprise systems in the post-adoptive phase, will offer new insights to the enterprise systems research and enables practitioners to consider their own practices.

The paper will first review the existing literature on enterprise systems and positions the study in relation to existing enterprise system literature. Next section reviews the literature on IS alignment with a special focus on enterprise systems. Third section reviews the literature on the post-adoptive stage of information systems. Section four develops the theoretical research model and discusses the research propositions. Finally, the paper presents some methodological issues and conclusions.

## 2 Enterprise systems and alignment

Enterprise systems are firmly entrenched in most of the firms today. Capable of supporting a majority of business processes, transactions, workflows, reporting and organizational structure, enterprise systems are considered a critical component of the current IT infrastructure in firms (Davenport et al 2004). By collecting greater amounts and types of internal data, enforcing business processes and controls, restricting and monitoring employee tasks, and supporting internal controls to a greater extent than ever before, ERP systems are changing the very nature of business (Grabski et al, 2011). Some of the capabilities delivered by the ERP systems have resulted in increased access to information, improved quality of information for decision making, and consistent and effective execution of business processes (Seddon et al, 2010). Studies, however have shown some ambiguous results. While some businesses have achieved operational efficiencies, others are struggling with to match the expectations and potential benefits of enterprise system deployment. Especially, the goals of improved performance in general, and agility and innovation in particular, as envisaged by the ERP software vendors, however, have not necessarily been realised (Galliers, 2007; Nazir & Pinsonneault, 2012). Especially, usage of the enterprise system's features and functionality and aligning and managing the organizational and managerial capabilities are some of the challenges organizations face.

Despite the accumulated knowledge about ERP projects, research on post-implementation effects of ERP systems on organisations is still limited (Peng & Nunes, 2009; Nazir & Pinsonneault, 2012). While there is a rich body of literature on ERP adoption and implementation, there is limited research on post-implementation effects and benefits such as flexibility, agility, process innovation, alignment and competitive advantage (Seddon et al 2010, Liang et al 2007). Given the investments on ERP systems, increasing costs for ongoing maintenance and upgrades, and the significant risk of failure, it is important for firms to understand the impact of enterprise systems use in the current dynamic business environment. Depending upon the varying contexts within which enterprise systems are situated and the changes they enable over time (Schubert & Williams, 2009), the organizational ability to use and exploit its full functionality and features could be different for different organizations. Staehr (2010) investigated how managerial agency influences the business benefits from ERP in the post-implementation stage. The study provided a better understanding of how managerial agency either enabled or constrained the achievement of benefits from ERP. That is why it is important to research managers' cognitive processes and actions in the post-adoptive stage.

Despite many years of research, alignment of information technologies/systems with business is one of the main concerns of management and has been ranked as the most important issue faced by organizations consistently since 1980s (Benbya and McKelvey 2006). Past studies identified various alignment levels, and analysed their impact on organizational performance, competitive advantage and processes, and on the integration of IS and business planning, domains and strategies (Tallon et al 2000). Though these studies have contributed to improved understanding of this phenomenon, their focus appears to be on 'simple cause-effect logic and ignored the multi-faceted and coevolutionary nature of IS alignment' (Benbya and McKelevy 2006). Annual CIO surveys conducted by the Society for Information Management consistently places this 'alignment' as a top priority (Preston 2014).

Alignment is defined as the 'degree to which the information technology mission, objectives and plans are supported by the business mission, objectives and plans (Reich and Benbasat, 1996:56). Alignment, in the context of enterprise systems (ES) adoption and use, is a continuous evolutionary process and continually reconciles the continuous discovery and adaptation of the ES features and functionality with the dynamic changes in the organizational capabilities, needs, priorities and environments.IT alignment, in spite of 30 years of research, still is an area of concern for many business organizations. It suggests that 'alignment is difficult and that it is a moving target" (Coltman et al 2015). It is therefore necessary to take a fresh approach to study IS alignment in the context of enterprise systems that are pervasive in most of the organizations. In contrast, tight and inflexible links between business and IT can also affect organizational agility and hamper performance (Tallon and Pinsonneault 2011, Seethamraju 2013).

In the traditional IS alignment literature, IS artefact once it is implemented, is taken for granted and considered unproblematic (Orlikowski and Iacono 2001). However, different components of alignment (IS/IT artefact and organizational components at individual, strategic and operational levels) are in a continuous and dynamic interaction with one another in a changing competitive environment. Thus, the alignment in this study is considered to be an emergent process of dynamic interactions between the three levels of analysis – strategic, operational and individual and the IS artefact (in this case ES software) with the need for them to be adjusted to sustain alignment. This co-evolutionary perspective allows the above process of mutual adaptation and change to be a dynamic interplay of coevolving interactions, interrelationships and effects (Benbya and McKelvey 2006).

There is a growing interest on the post-implementation phase of enterprise systems particularly so on the sociological aspects, interoperability and optimization of the system in organization and on the benefits management (Staehr 2010). A study by Staehr (2010) using the lens of the structuration theory (Giddens, 1984), investigated the role of managerial agency in ERP projects. Besides Staehr's research the literature focusing on managerial action in this context is scarce and there is a need for a better understanding of the managers' cognitive processes and action in the post-adoptive stage of ERP systems. The existing information systems (IS) literature, including that of IS evaluation and benefits management, recognize the need to continuously evaluate IS and to identify further benefits of IS, but they do not provide much further understanding of how this actually happens in organisations. Recent research on benefits management capability (Ashurst et al., 2008) identifies practices ensuring fully utilizing the potential of IS as an organizational capability. While these researchers refer to concrete practices they are silent about what people actually do when enacting these practices. There is thus a clear gap in the existing literature in understanding how managers consider the full utilization of ERP systems in the post-adoptive stage.

## 3   Post-adoptive Stage of Information Systems

The aim of this section is to give a brief overview of IS literature that has addressed issues on the post-adoptive stage of information systems. It provides the necessary background for the research model that will be introduced in section four and helps to understand the contribution of the present paper.

### 3.1   IS Evaluation and Benefits

The existing frameworks and models on IS evaluation and benefits management help to assess investment alternatives, broadly describe ideas for on-going evaluation and describe steps for benefits management. However, they are not able to explain how new opportunities for utilizing enterprise systems (or information technology in general) are discovered in an organisational context.

Formal IS evaluation methods are frequently considered being of little value by practitioners (Jones and Hughes, 2001). More than 60 IS evaluation methods have been proposed in the literature (Renkema and Berghout, 1997), but in practice only the simple ones are applied. It was generally accepted that IS evaluation should take place in all phases of the system life-cycle (Willcocks and Lester, 1993) and some researchers suggested that evaluation should be a continuous activity (Remenyi and Sherwood-Smith, 1999). It is rather evident that the research on IS evaluation has mainly focused on ex ante evaluation and on developing broad conceptual frameworks and procedures for evaluation.

The existing literature on IS evaluation is silent about the process by which decisions on improvements, re-configurations or updates of enterprise systems are made in organisations.

Researchers started to see IS evaluation as a complex organisational activity and suggested interpretive methods for IS evaluation (Stockdale and Standing, 2006). The interpretive stream of IS evaluation research recognized that evaluation is situated and contextual. This has led to a call for more interpretive, local, and "situated" evaluation procedures (Jones and Hughes, 2001; Stockdale and Standing, 2006). Against the backdrop of the non-use of the formal evaluation methods, re-directing IS evaluation research to more interpretive approaches seems promising; however, the existing literature is not able to provide an understanding what managers actually do when they evaluate IS.

The literature on benefits management identifies a step for seeking further benefits in the post-adoptive stage (Ward and Daniel, 2012), but the existing frameworks provide little understanding on how this actually happens in an organisational context. The existing research results show that companies do not often apply the benefits management approach or follow its principles. Researchers have investigated both the extent of the adoption of the benefits management principles and the factors affecting adoption of these principles (Lin and Pervan, 2003; Paivarinta et al., 2007). Many researchers have recognised the importance of understanding IT-related change and have proposed models to provide a better understanding of this (Ward and Elvin, 1999; Staehr, 2010).

Recently, a growing body of literature examining the so called 'benefits management capability' has emerged (Ashurst et al., 2008; Ashurst and Hodges, 2010). Ashurst et al. (2008). They have identified the following capability for the benefits exploitation phase (post-adoptive phase), defined as 'the adoption of the portfolio of practices required to realize the potential benefits from information, applications and IT services, over their operational life' (Ashurst et al. 2008, p. 356). While they point to concrete practices, a more detailed understanding is needed to understand what managers actually do when ensuring that the full potential of information systems (IS) is utilized.

### 3.2 Post-adoptive Behaviour of IS Users

The post-adoptive behaviour of users has recently received more attention from researchers. Jasperson et al. (2005) presented a two-level model of post-adoptive behaviour; their model includes both the cognitive processes of technology users and the work system interventions on the organisational level. Reflective cognitive processing, technology sensemaking, may affect individuals' post-adoptive intentions thus affecting future post-adoptive behaviour. Similarly, users engage in work system sensemaking regarding work system outcome expectation gaps. Importantly, these gaps may be perceived differently by different groups such as users, technology experts and managers (Jasperson et al., 2005).

Another stream of research which can inform us about post-adoptive behaviour is research applying the concept of affordance (Gibson, 1986; Strong et al., 2014). Drawing from Gibson's (1986) original theory Strong et al. (2014) presented an extended definition of affordances as follows: "the potential for behaviours associated with achieving an immediate concrete outcome and arising from the relation between an artefact and a goal-oriented actor or actors" (p. 69). The concrete outcomes of enterprise systems use are thus determined by the interaction between the users and the features of the enterprise system.

The above theories, while valuable, have a rather strong focus on the IT artefact. However, information systems are used in a context and as part of (social) practices. As Riemer and Johnston (2014) put it: "The entity encountered is not defined by its properties, but first and foremost by its place in a practice that makes intelligible the object and influences which properties show up for us as meaningful in a situation" (p. 279).

### 3.3 External Influences on Enterprise Systems Alignment Decisions

The decisions regarding the adoption or upgrades of ERP systems may be affected by external influences. Managers are likely to communicate with ERP system vendors, the implementation partners and representatives of other companies in the industry when considering ERP solutions.

Benders et al. (2006) investigated the isomorphic pressures related to ERP adoption and introduced the term 'technical isomorphism', which "manifests itself in the enactment of blueprints for centralization and standard working procedures that are embedded in the ERP-software" (Benders et al., p. 194). This effectively leads to the adaptation of the organisation to the system. Researching outsourcing decisions Blaskovich and Mintchik (2011) found that senior accounting executives often mimic the outsourcing actions of their industry peers, particularly if they perceive the skills of their chief information officer as weak. Chang et al. (2008) showed that external communication channels were highly important in the diffusion of ERP technology in Taiwan. Das and Nair (2010) found that the adoption of specific manufacturing technologies was affected by the specific supply chain management practice in use. Although the above examples of research on the external influence towards IS decisions mainly concern the technology adoption decisions it can be assumed that similar external pressures can be present in making decisions on ERP solutions in the post-adoptive stage as well.

# 4  Research Model and Research Questions

## 4.1  The original model of creative search and strategic sense-making

Sensemaking directs attention to organising and action, i.e. what people actually do. In doing so, it helps to create a more detailed understanding of how actors make sense of situations and how they enact remedies to issues. Sensemaking is triggered by anything that causes a need for reflection, such as an issue with the use of an enterprise system or a general business issue.

Sensemaking is heavily based on identity beliefs. It is triggered by discrepancies and employs noticing and bracketing to make sense of the situation. Because people have different identities and, in particular in organisations, different roles they notice (pay attention to) different cues in the environment. Based on these cues they create a plausible story of the situation and enact solutions with the aim to create stability. Plausibility is the fundamental criterion for sensemaking, rather than accuracy. (Weick, 1995)

Pandza and Thorpe (2009) advocate the role of managerial agency for creating dynamic capabilities in organisations. They argue that, while knowledge may accumulate through experiential learning, "significant deviations in knowledge progression occur by purposeful and creative managerial engagement with existing patterns and routines that shape this progression of knowledge" (Pandza and Thorpe, 2009, p. 128).

Pandza and Thorpe (2009) define strategic sensemaking "as an uncertainty-reducing cognitive process of initial sensemaking that activates purposeful action and retrospective sensemaking that enables managers to understand the appropriateness and usefulness of the development of novel knowledge and its fit into business opportunities" (p. 124). The sequence starts with initial sensemaking which is "seen as a cognitive process that by interpreting information derived from the external environment and from internal learning instigates purposeful action that leads to change" (Pandza and Thorpe, 2009, p. 123). The next step is the search activity where individuals search for feasible alternatives that would satisfy certain defined criteria. Imagination and intuition are part of the search because the activity concerns the future. Pandza and Thorpe argue that initial sensemaking does not necessarily initiate the search activity for new opportunities, but may also result in extending the existing knowledge trajectories. Initial sensemaking and creative search are complementary cognitive processes; neither of them alone is sufficient to develop new knowledge (Pandza and Thorpe, 2009).

The search process is purposeful and ranges from searching alternatives to the recognition of opportunities and to further exploration of the alternatives. An opportunity (and the consequences of it) needs to be aligned with the initial sensemaking in order to initiate further exploration. Sensegiving engages other organisational members and disseminates the new understanding in the organisation (Foldy et al., 2008). Retrospective sensemaking "provides interpretation of the usefulness of the newly developed knowledge" (Pandza and Thorpe, 2009, p. 124), thus reducing uncertainty and creating stability (Weick, 1995).

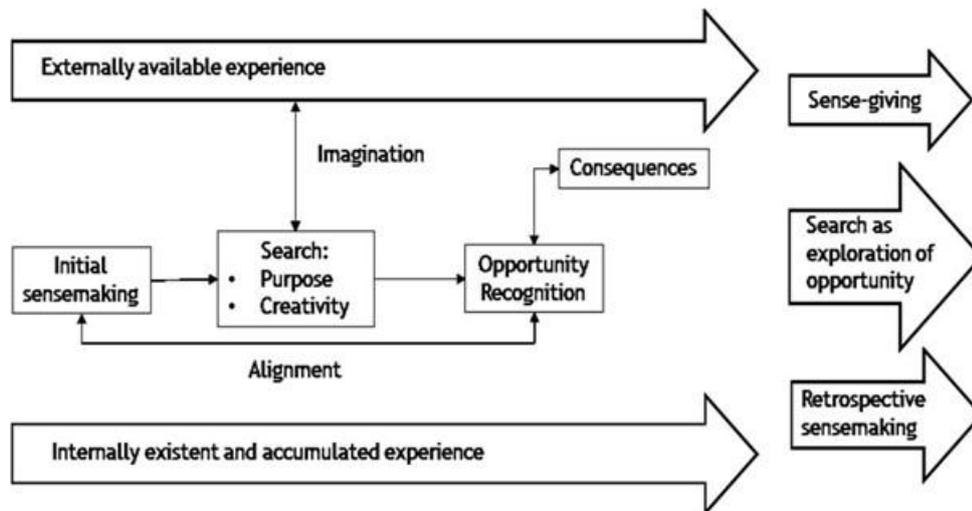

*Figure 1. The model for creative search and strategic sensemaking (adapted from Pandza and Thorpe, 2009)*

### 4.2  The Proposed Research Model and Research Questions

Our proposed research model (Figure 2) focuses on the cognitive processes and action leading to decisions regarding enterprise systems in the post-adoptive phase. The model focuses on understanding how managers determine what are the feasible solutions and what criteria they apply for choosing a solution. Because an individual manager has his/her individual work history and experiences as well as a certain organisational role and identity, they would pay attention and notice different factors when initially making sense of the situation and the business problem at hand. Their prior experience and professional identity would also influence how they would define the criteria that a solution would have to satisfy. Because of their different professional background, the views of e.g. Chief Information Officer, Chief Technology Officer and/or Chief Finance Officer are likely to be rather different representing the IS, IT and Finance/accounting views of the business problem respectively. Identifying these differences and their influence on successful solutions will be central to our empirical work. The initial view of a feasible solution and the criteria then guides a purposive search process where suitable solutions are identified and considered. Following Pandza and Thorpe (2009), imagination is part of the search process because it is essentially about a future solution. External and internal available knowledge will influence the manager's perception about the future solution and there is likely to be both internal and external pressures, such as from user groups or industry partners, which will affect the perception of the feasible solution. In our model we want to emphasize the consequences of committing to a certain solution, which effectively creates constraints for the organisation. These constraints will only be visible when the new solution is in use and it will only at this stage become clear whether the business problem could be resolved. Constraints related to the technical or process aspects of a solution may indeed create new business problems, which initiates the creative search process again.

In our model, we address the creative search process from the alignment perspective. Based on the existing IS literature we are able to identify broad categories of feasible alternatives that a manager would likely consider. These categories are not meant to be definite at this stage, but rather they are meant to provide guidance for empirical work and form initial broad categories to help coding empirical data. They represent our initial understanding based on literature and we will be open for other categories to emerge while conducting empirical research. Based on the existing IS literature, we suggest that there are four different feasible alternatives for alignment: 1) changing (improving, configuring, updating) the current enterprise system 2) changing the organisational capabilities (knowledge, skills, processes, practices) 3) sourcing solutions from outside of the organisation (new ES, components, add-ons) 4) aligning with or affecting the external environment of the organisation (by e.g. influencing the supply chain partners). Based on the goal of functional integration (Henderson and Venkatraman, 1993) a feasible solution for achieving alignment could be either changing the enterprise system (option 1) or changing the organisational capabilities (option 2). Sourcing solutions from outside could be considered as a feasible solution because of isomorphic pressures (Benders et al., 2006) and if a standard solution is considered feasible or it is considered necessary to utilize knowledge and expertise from outside of the company (Saarinen and Vepsalainen, 1994). Aligning with or affecting the external environment could be considered as a feasible option, for example, because of specific supply chain management practices (Das and Nair, 2010).

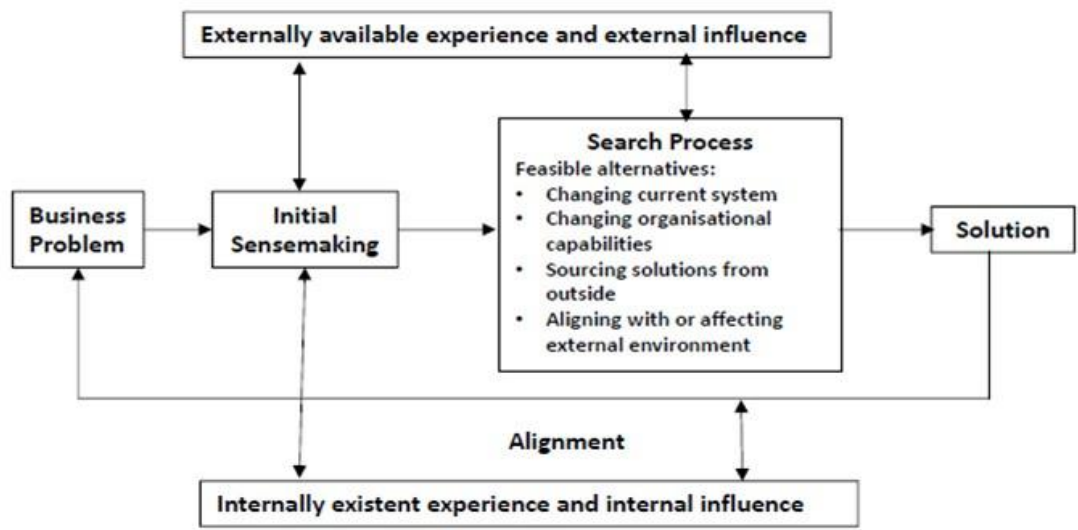

*Figure 2. The research model for the present study*

Below we elaborate on the factors affecting managerial considerations about the feasible solutions and the criteria they are likely to apply when considering the solutions.

**Changing the current system.** This involves adjusting and/or modifying the system and technology for a solution, including customizing and/or configuring the existing software solution, and/or going for an upgraded version. Many a times, it is possible to find a solution from the existing system. It is possible to solve a business problem by customizing and/or configuring the existing system and/or moving to an upgraded version with better features and functionality. Ability of organizations to align its enterprise system to changing business requirements depends upon the modular flexibility. As identified by Nelson et al (1997), there are two aspects of modular flexibility – structural flexibility that refers to the characteristics such as modularity and acceptance of change, while process flexibility refers to people's ability to change the technology. Modularity in the enterprise systems context, is the ability to easily reconfigure (add, modify, or remove) system components by minimizing the interdependencies among modules. In case of enterprise system, it means further customization of the software solution and/or configuring additional features and functionality to meet business requirements.

**Changing organisational capabilities.** As enterprise systems are complex information systems capable of creating efficiencies through process integration, standardization and real-time visibility, it is necessary for organizations to find ways to assimilate and exploit the systems to meet ever changing business requirements. Organization's ability to achieve benefits of enterprise systems is dependent upon recognizing, exploring and using the enterprise system features and functionality. ES usage refers to how users employ the system features to perform a task (Burton-Jones & Gallivan 2007, Nwankpa & Roumani, 2014). A feasible alternative to improve organisational capabilities could be to engage users in reflecting on their work practices and their use of the enterprise system functionality. Other feasible alternatives may involve user skills development or managerial action to influence the cultural aspects in the organisation.

**Sourcing solutions from outside.** Purchasing a new ES, new components or add-ons could be considered as a feasible alternative for various business or technological reasons e.g. when the existing ES is out of date or a company wants to adopt a clean slate approach to renew its enterprise systems, or a company wants to adopt best practice business models (Hallikainen et al., 2004). These decisions may be affected by the isomorphic pressures as mentioned above (Benders et al., 2006). However, information systems in general and enterprise systems that have endured significant change management across the organization, typically are more disablers of flexibility rather than enablers (Allen and Boynton 1991). Thus, the consequences of implementing a new ES must be considered. Most of the enterprise systems implementations generally automate the status quo with a few process improvements here and there, and generally result in resistance to change and inertia (Seddon et al 2010).

**Aligning with or affecting the external environment.** Here the issues related to the supply chain are important in the case of ES. A company may have to align its ES with the dominant supply chain practices (Das and Nair, 2010) if this was the cause for the business problem identified. Depending on the organisation's position, it may also be possible to affect the supply chain partners to adjust their technological solutions.

This sensemaking process in general and the search process in enterprise systems alignment context in particular are heavily based on identity beliefs. Due to the differences in the roles played by people, their professional orientation and the consequent attention paid to different cues in the environment, the solutions enacted to align the enterprise systems with business requirements are different. Rather than accuracy, plausibility is the main criterion in identifying and prioritizing the feasible alternatives. The proposed framework therefore gives rise to two research themes. Firstly, it is important to understand how decision makers in different managerial roles perceive a solution as feasible and how they make decisions on the solutions to be adopted and how they deal with the consequences. Secondly, from the organisational efficiency and effectiveness perspective, it is important to consider the consequences of these decisions and whether they lead to successful solutions in organisations and how these managers deal with those consequences. Thus, we propose the following two broad research questions for our study:

How does a manager's organisational role, level and experience affect their views on and the development and implementation of the feasible alternatives in aligning the ES with business requirements?

What are the consequences of implementing certain solutions, how successful are they and how do the solutions evolve over time?

Answering the research questions above will provide new knowledge on the organisational decision making related to ES as well as on the dynamics between different solutions and emerging business needs. The research methodology is explained in the next section.

## 5 Methodology

Recognising its innovative nature of investigating the alignment as a result of the co-evolution of the enterprise system's features, business changes and users' behaviour; and the nature of research questions, a mixed method approach will be adopted as explained below.

Even though enterprise systems are in use for more than a decade now, empirical data on the way alignment is taking place and how organizational capabilities, ES use and ES features and functionality evolve over time in business organizations were limited. Further, we do not know the implications of this coevolution for benefits realisation and organizational performance. Therefore, our first phase of study will employ a multi-case research method in order to understand the phenomenon at organisational as well as at individual levels and takes a multiple stakeholders perspective. The objective is to understand and explore the antecedents, barriers, inhibitors and factors influencing the continuous process of aligning ES with changing business requirements within the organisational context taking into consideration different types of business changes, organisational and contingent variables. The aim here is to explore the phenomenon from three aspects – technology (ES software related), organizational (capabilities including skills, resources for training and customization, priorities, individual managerial preferences, work-arounds and changing business needs) and environmental factors (such as competition, software vendors pressure to upgrade, level of support from the software vendor, suppliers, customers) and other contingencies in each organizational context. In addition, this phase will document the alignment phenomenon in terms of the identification of alternative solutions, evaluation of benefits and benefits realisation, adaptation of the ES software solutions to realise the intended benefits, changes to organizational capabilities (both positive and negative) from the individual managers' perspective taking into consideration their background, professional orientation and beliefs.

Case study organizations in this phase will be purposefully identified and organizations that have an enterprise system in place at least for a period of eight years will be selected. Data will be collected primarily through semi-structured interviews. The objective is to collect information about five key aspects: i) scope, types, nature and extent of usage of various ES features configured and/or customized in the organization, ii) the antecedents for the identification, evaluation, adoption or non-adoption of feasible alternative solutions including iii) the factors influencing the managerial attitude, beliefs and preferences for various alternatives in terms of the extent of use, 'work-arounds', skills, roles, work practices, and their impact or consequences and, v) impact of the co-evolution of these three aspects (system features, system use and business changes) from individual manager's perspective in specific organisational contexts over time. The focus at individual managerial level will be on their attitude, beliefs and perceptions in terms of changes and/or evolution of the ES and business needs/problems and the associated business changes experienced by the users.

All the semi-structured interviews will be transcribed and the interview transcripts and related observational notes will be sent to individual respondents for validation and corroboration. The validated data will then be analysed by cataloguing the text according to interview transcripts, observational notes and internal documentation. In addition, organisational case and relevant demographic data (position, seniority, age, gender, professional association, qualifications, experience etc.) of the respondents will also be collected. The coding would categorise the text not only in terms of its empirical description, but also based on the strength and textual role, for example by asking questions such as: Is the text intended to identify, reassure and create confidence in the behaviour of users and use of ES features? Does it promote a sense of urgency to make appropriate business and behavioural changes and/or customization of the ES features and functionality or a sense of complacency? Does it seek to justify their use behaviour, business changes and co-evolution, or explain, deny or criticise?

Appropriate validation efforts such as checking for research effects and biases, providing interview manuscripts to respondents, getting feedback from respondents and triangulation of data will be made to improve the data quality. The data will be analysed in an iterative manner with data collection and emerging themes and propositions/hypotheses will be identified.

In the second phase of the study, a questionnaire survey will be conducted to measure the alignment processes and alternatives, and changes in the benefits realised and consequences. Further, it will measure the extent of changes in all the three constructs – ES use, ES features and business changes, the factors influencing the co-evolution, effective alignment and the barriers for the realization of alignment benefits. This survey will adopt multiple-respondent strategy and will obtain 3 to 6 responses from each of the organization including business functional managers responsible for the use, IT/IS executives/managers responsible for the customization and deployment of the features and senior managers authorising and/or directing the selection and implementation of alternatives and dealing with their consequences The data thus collected will be analysed using standard statistical methods testing various hypotheses developed from the previous multiple case study research phase.

This study has limitations and the generalisability of the findings are limited given that the focus is on sense making influenced by the individual identity beliefs and organizational contexts. Similar to any other qualitative study, our first phase of the study, may project the subjective bias of the researcher and the respondents and could lack generalisability. Further, the organizations to be studied are not randomly chosen and therefore it is possible that the organizations that would participate in this study may have a success story to share rather than challenges they might have faced and the options they might have considered. As sensemaking is based on individual identity beliefs and organizational contexts, it could be influenced by the unique phenomena in each organization and individual respondents may overrate or underrate the issues. The second phase of the study, by measuring the changes in the adaptation and alignment of all the three constructs – ES use, ES features and 'business changes', would offer opportunities for generalisability of the findings. It too, however, may be subject to typical limitations of a quantitative survey.

# 6   Conclusion

Aligning enterprise systems with the dynamic business requirements is a complex and continuous phenomenon and individual managerial beliefs and identity play an important role in shaping, prioritizing and adopting the alternative solutions. This paper by adapting the creative search process in the sensemaking model of Pandza and Thorpe (2009) from alignment perspective in the post-adoptive ES context, contributes to the practice theory (Nicolini 2013). By focusing on understanding what managers actually do and how they interpret what they are supposed to do, this research model helps to conduct research that is practically more relevant. Though they describe the relations between managerial attitudes and beliefs, existing models on IS evaluation and benefits management fail to increase our understanding of what actually happens in terms of managerial cognition and action. Understanding what managers do will help researchers to consider the implications of the managerial action and potentially to identify root causes for ES related problems in organisations. It guides a reflective manager to consider the organisational issues related to ES decisions.

# 7　References